\documentclass[prl,
                preprint,
                amsmath,
                amssymb,
                showpacs,
                superscriptaddress]{revtex4-1}
\usepackage{graphicx} 
\usepackage{dcolumn}  
\usepackage{bm}       
\usepackage{amsfonts}
\usepackage{dsfont}
\usepackage{ulem}
\usepackage{color}

\usepackage{ulem}
\usepackage{color}
 \newcommand{\Madd}[1]{\textcolor{blue}{#1}}
 
\begin{document}

\title{High-Harmonic generation from spin-polarised defects in solids}

\author{Mrudul M. S.}
\affiliation{%
Department of Physics, Indian Institute of Technology Bombay,
            Powai, Mumbai 400076, India }

\author{Nicolas Tancogne-Dejean}
\affiliation{%
Max Planck Institute for the Structure and Dynamics of Matter, 
Luruper Chaussee 149, 22761 Hamburg, Germany  }

\author{ Angel Rubio}
\email[]{angel.rubio@mpsd.mpg.de}
\affiliation{%
Max Planck Institute for the Structure and Dynamics of Matter, 
Luruper Chaussee 149, 22761 Hamburg, Germany  }
\affiliation{%
Nano-Bio Spectroscopy Group and ETSF,  Departamento de Fisica de Materiales, Universidad del Pa\'is Vasco UPV/EHU- 20018 San Sebasti\'an, Spain }
                        
\author{Gopal Dixit}
\email[]{gdixit@phy.iitb.ac.in}
\affiliation{%
Department of Physics, Indian Institute of Technology Bombay,
            Powai, Mumbai 400076, India }

\date{\today}


\begin{abstract}
The generation of high-order harmonics in gases enabled to probe the attosecond electron 
dynamics in atoms and molecules with unprecedented resolution.
Extending these techniques to solids, 
which were originally developed for atomic and molecular gases, 
requires a fundamental understanding of the physics  
that has been partially addressed theoretically. 
Here, we employ time-dependent density-functional theory to investigate how the electron 
dynamics resulting in high-harmonic emission in monolayer hexagonal boron nitride is affected 
by the presence of vacancies. We show how these realistic spin-polarised defects modify the 
harmonic emission and demonstrate that important differences exist between harmonics from 
a pristine solid and a defected solid. In particular, we found that the different spin channels are 
affected differently by the presence of the spin-polarised point defect. Moreover, the localisation 
of the wavefunction, the geometry of the defect, and the electron-electron interaction are all 
crucial ingredients to describe high-harmonic generation in defected solids. 
\end{abstract}

\maketitle 
\section{Introduction}

Recent advances in mid-infrared and terahertz laser sources have demonstrated the 
generation of non-perturbative high-order harmonics from solids, including semiconductors, 
dielectrics,  and nano-structures below their damage threshold~\cite{ghimire2019, 
ghimire2011observation, ghimire2011redshift, zaks2012experimental, schubert2014sub, 
vampa2015all, vampa2015linking, hohenleutner2015real, luu2015extreme, ndabashimiye2016solid, you2017high, 
lanin2017mapping, sivis2017tailored, langer2018lightwave}.  With the pioneering work of 
Ghimire et al.~\cite{ghimire2011observation}, high-harmonic generation (HHG) in solids   
offers fascinating  avenues to probe light-driven electron dynamics in solids on attosecond 
timescale~\cite{silva2018high, silva2018all, bauer2018high, chacon2018observing, 
reimann2018subcycle, floss2018ab}, and to image energy band-dispersion of 
solids~\cite{vampa2015all, ndabashimiye2016solid, lanin_mapping_2017}.
Moreover, due to the high electron density in solids in comparison to gases, HHG from solids 
may be superior for higher harmonic yield. 
Finally, HHG in solids represents an attractive route towards compact table-top light-source for 
coherent and bright attosecond pulses in the extreme ultraviolet and soft x-ray energy 
regime~\cite{ghimire2019, luu2015extreme,  vampa2017merge, kruchinin2018colloquium}.

Due to the growth processes, defects are inevitable in real solids~\citep{hayes2012defects}. 
Defects in materials can  appear in the form of vacancies, impurities, interstitials (all of these 
can be neutral as well as charged), dislocations, etc.
Defect-induced microscopic modifications in a material 
significantly affect on its macroscopic properties~\cite{wilson1931theory}.  
Electronic, optical, vibrational, structural and diffusion  properties of solids with defects have 
been thoroughly reviewed over the past century~\cite{
barker1975optical, pantelides1978electronic, queisser1998defects, van2004first, bockstedte2010many, 
alkauskas2011advanced, freysoldt2014first}. 
Defect engineering is 
used  to achieve desirable characteristics for materials, e.g.,  doping 
has revolutionised the field of electronics~\cite{bardeen1949physical}. 
Defects can also be highly controlled\Madd{,} and it is possible to create isolated defects such 
as nitrogen-vacancy defects  in diamond~\cite{doherty2013nitrogen, maze2008nanoscale} or 
single photon emitters in two-dimensional materials~\cite{tran2016quantum}.

The  influence of defects in solids  is not well explored in  strong-field physics. 
In this work, we aim to address the following questions:  
Is it possible to observe defect-specific fingerprints in strong-field driven electron dynamics?
Or does the electron-electron interaction play a different role for defects and bulk materials?   
Moreover, some defects are also spin polarised in nature. This raises the question if it is 
possible to control the electron dynamics for different spin channels independently in a non-
magnetic host material. As we will show below, HHG is a unique probe that helps us to explore 
these interesting questions. 

We aim to model strong-field driven electron dynamics in a defected-solid with the least 
approximations. In this work, 
we employ \textit{ab initio} time-dependent density functional theory (TDDFT) to simulate the 
strong-field driven electron dynamics in defected-solid~\cite{tancogne2017impact}. This allows 
us to come up with theoretical predictions relevant for real experiments on defected-solid 
without empirical input.
There are various theoretical predictions for the HHG from doped semiconductors by
using simple one-dimensional  model Hamiltonians~\cite{yu2019enhanced, huang2017high, pattanayak2020}. 
In Ref.~\cite{yu2019enhanced}, 
by using TDDFT, it is predicted  that there is an enhancement of several orders of magnitude in 
the efficiency of HHG in a donor-doped semiconductor.
Using an independent particle model, Huang et al. contrastingly found that 
the efficiency of the second plateau from the doped semiconductor is 
enhanced~\cite{huang2017high}. 
Similar calculations within a tight-binding Anderson model indicates that the disorder 
may lead to well-resolved peaks in  HHG~\cite{orlando2018high}. 
The conceptual idea for the tomographic imaging  of shallow impurities in solids, within a one-dimensional hydrogenic model, has been developed by Corkum and co-workers~\cite{almalki2018high}. 
Even though these pioneering works have shown that defects can influence HHG, many points 
remain elusive. So far, only model systems in one dimension have been considered, and no 
investigation of realistic defects (through geometry optimisation and relaxation of atomic forces) 
has been carried out. 
Beyond the structural aspect, several other essential aspects need to be investigated in order 
to obtain a better understanding of HHG in defected-solids such as the importance of electron-electron interaction (that goes beyond single-active electron and independent-particle  
approximations),  the role of the electron's spin, the effect of the symmetry breaking due to the 
defects, etc. 
Our present work aims to shed some light on some of these crucial questions. For that, 
we need to  go beyond  the one-dimensional model Hamiltonians. 
  
In order to investigate how  the presence of defects modifies HHG in periodic materials, we 
need to select some systems of interest. Nowadays, two-dimensional 
(2D) materials are at the centre of tremendous research activities as they 
reveal different electronic and optical properties compared to the bulk solids. 
Monolayer 2D materials such as transition-metal dichalcogenides~\cite{liu2017high, 
langer2018lightwave}, graphene~\cite{yoshikawa2017high, al2014high} and hexagonal boron 
nitride (h-BN)~\cite{tancogne2018atomic,  le2018high,yu2018two} among others have been 
used to generate strong-field driven high-order harmonics.  
Many studies have examined HHG in h-BN when the  polarisation of the laser pulse is either in-
plane~\cite{le2018high, yu2018two} or out-of-plane~\cite{tancogne2018atomic} 
of the material.  
Using an out-of-plane driving laser pulse, Tancogne-Dejean et al. have shown that atomic-like 
harmonics can be generated from h-BN~\cite{tancogne2018atomic}.  Also, h-BN  is used to 
explore the competition between atomic-like and bulk-like characteristics of  
HHG~\cite{tancogne2018atomic, le2018high}. In MoS$_2$, it has been demonstrated 
experimentally  that the generation of  high-order harmonics is more efficient in a monolayer in 
comparison  to its bulk counterpart~\cite{liu2017high}.  
Moreover,  HHG from  graphene exhibits an unusual dependence on the laser 
ellipticity~\cite{yoshikawa2017high}. 
Light-driven control over the valley pseudospin in WSe$_2$ is demonstrated by 
Langer et al.~\cite{langer2018lightwave}. 
These works have shed light on the fact that 2D materials are promising   
for  studying light-driven electron dynamics and for more technological applications in petahertz 
electronics~\cite{garg2016multi} and valleytronics~\cite{schaibley2016valleytronics}. 

Monolayer h-BN is an interesting material for the  study of electronic and  optical properties. 
h-BN is a promising candidate for light-emitting devices in the far UV region due to the strong 
exciton emission~\cite{bourrellier2014nanometric, bourrellier2016bright}. 
Due to this technological importance, 
several experimental and theoretical studies have been carried out for h-BN with 
defects~\cite{tran2016quantum, tran2016robust, huang2012defect, zobelli2007vacancy, alem2009atomically, jimenez1996near, suenaga2012core, liu2014direct, thomas2015temperature, gilbert2017fabrication, alem2009atomically, alem2011probing,  wong2015characterization, pierret2014excitonic,
bourrellier2016bright, attaccalite2011coupling, azevedo2009electronic, orellana2001stability, mosuang2002influence}. Different kinds of defects in h-BN can be classified as vacancy 
(mono-vacancies to cluster of vacancies), antisite, and impurities. In particular, defects like 
monovacancies of boron and nitrogen atoms in h-BN are among the most commonly observed 
defects.  Recently,  Zettili and co-workers have shown the possibility of engineering a cluster of 
vacancies and characterising them using ultra-high-resolution transmission electron 
microscopy~\cite{gilbert2017fabrication, alem2009atomically, wong2015characterization}. 
Signatures of defects in h-BN are identified by  analysing    cathodoluminescence and 
photoluminescence spectra ~\cite{pierret2014excitonic, bourrellier2014nanometric}. It is shown 
that the emission band around 4 eV originates from the transitions involving deep defect 
levels~\cite{attaccalite2011coupling}. Ultra-bright single-photon emission from a single layer of 
h-BN with nitrogen vacancy is achieved experimentally, for example in large-scale nano-photonics and quantum information processing~\cite{tran2016quantum, tran2016robust}. 
In Refs.~\cite{huang2012defect, attaccalite2011coupling, azevedo2009electronic,
liu2007ab, orellana2001stability}, 
different kinds of defects in h-BN are modeled, and their effects on the electronic and optical 
properties are thoroughly investigated.  

In the present work, 
using the well-established supercell  approach to model the defects in h-BN, 
we analyse their influence on HHG. Due  to the partial ionic nature of its bonds, h-BN is a wide 
band-gap semiconductor with an experimental band-gap of 6 eV. 
This  makes h-BN an  interesting candidate for generating high-order harmonics without 
damaging the material. Moreover, 2D material like h-BN enable us to easily visualise the 
induced electron density and localised defect states easily. The presence of boron or nitrogen 
vacancies in h-BN acts as a spin-polarised defect. These factors make h-BN an ideal candidate 
for exploring spin-resolved HHG in non-magnetic defected solids. Below, we will discuss HHG 
in h-BN with and without spin-polarised monovacancies.

\section{Results}

\subsection{Computational approach}
h-BN has a  two-atoms primitive cell. We model the vacancy in a 5$\times$5 (7$\times$7) 
supercell with 50 atoms (98 atoms) (See methods section for the details of structural 
optimisation) following the methods in  Ref~\cite{huang2012defect}. The size of the 5$\times$5 
supercell is large enough to separate the nearest defects with a distance greater than 12 \AA. 
This minimises the interaction between the nearby defects~\cite{huang2012defect,liu2007ab}, 
and the defect wavefunctions are found to be well-localised within the 
supercell~\cite{azevedo2009electronic}. 
In the present work, 
we are not considering more than one point defect.  
Both of our vacancy configurations (single boron as well as single nitrogen vacancy) have a 
total magnetic moment of +1$\mu_B$, which is consistent with the earlier reported  results 
~\cite{huang2012defect,liu2007ab,azevedo2009electronic}. 
For the low defect-concentration considered here (2\% and $\sim1\%$), the strength of the total 
magnetic moment is independent of the defect concentrations~\cite{liu2007ab}. We have found 
that the 7$\times$7 and 5$\times$5 supercells converged to similar ground-states~\cite{huang2012defect}. 

In all the present calculations, we consider a laser pulse of 15-femtosecond duration at  full-
width half-maximum with sine-squared envelope and a peak intensity of $13.25$ TW/cm$^2$ in 
matter (for an experimental in-plane optical index $n$ of $\sim 2.65$~\cite{PhysRev.146.543}). 
The carrier wavelength of the pulse is 1600 nm, which corresponds to a photon energy of 0.77 
eV. The polarisation of the laser is linear and its direction is normal to the mirror plane of h-BN. 
The symmetry of the pristine h-BN permits the observation of  the harmonics in the parallel and 
perpendicular directions to the laser polarisation. The direction resolved analysis of the HHG
spectrum is shown in the Supplementary Information. 
For pristine h-BN, the band gap is found to be 4.73 eV within DFT-PBE level. The energy band-gap falls near the sixth harmonic of the incident photon energy and laser parameters are well-below the damage threshold of the pristine.

\subsection{High-harmonic generation in hexagonal boron nitride with a boron vacancy}

\begin{figure}[h!]
\includegraphics[width= 13 cm]{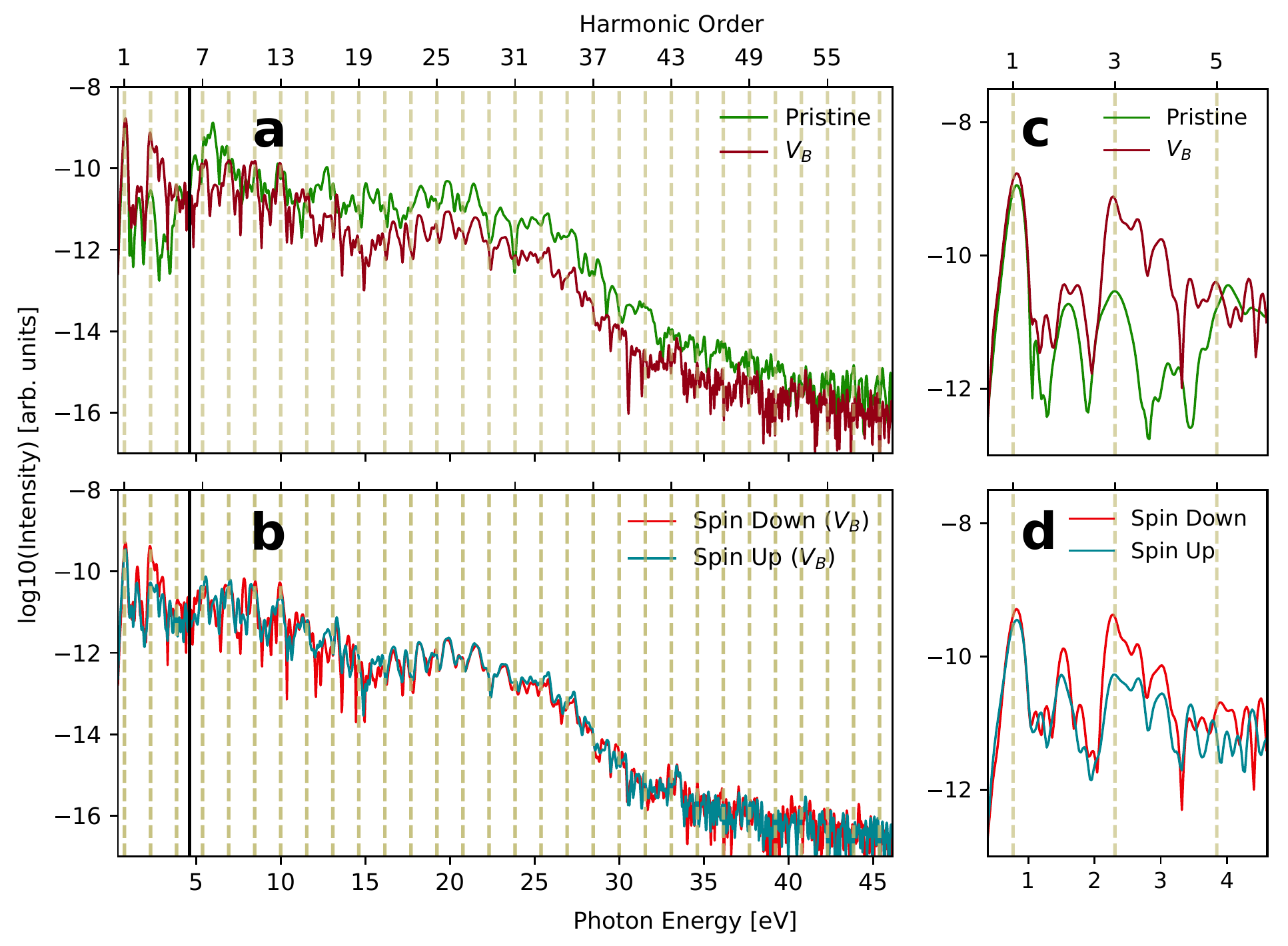}
\caption{{\bf High-order harmonic spectrum of monolayer hexagonal boron nitride (h-BN) with a boron vacancy}.   
{\bf{(a)}} The total harmonic spectrum of boron-vacant h-BN ($V_{B}$) and pristine h-BN. The black vertical line represents the energy band-gap of h-BN. {\bf{(b)}} Spin-resolved harmonic spectrum of $V_B$. {\bf{(c)}} and {\bf{(d)}} correspond, respectively, to the below band-gap harmonics of {\bf{(a)}} and {\bf{(b)}}.} 
\label{fig1}
\end{figure}

We start our analysis by comparing the HHG from pristine h-BN  and from 
h-BN with a boron vacancy. Removal of a boron atom makes the system spin-polarised~\cite{huang2012defect,liu2007ab}.  
The high-harmonic spectrum of  h-BN with a boron vacancy ($V_{B}$) and its 
comparison with the spectrum of pristine h-BN is presented in Fig.~\ref{fig1}a. 
The spectrum of $V_{B}$ is different from the pristine h-BN as evident from the figure. 
There are two distinct differences: First, the below band-gap harmonics corresponding 
to  $V_{B}$ are significantly enhanced (see Fig.~\ref{fig1}c). 
Second, the harmonics,  closer to the energy cutoff, have a much lower yield for $V_{B}$ 
in comparison to the pristine material. \\
An interesting aspect for the spin-polarised defects in a non-magnetic host is that the effect of 
the defect states, compared to bulk states, can be identified clearly by 
examining the spin-resolved spectrum. 
The harmonics corresponding to spin-up and spin-down channels in $V_B$ are 
shown in Fig.~\ref{fig1}b. As reflected in the figure, the below band-gap harmonics are different 
for both the channels. 
The strength of the third harmonic corresponding to the spin-down channel is much stronger 
(by an order of magnitude)
in comparison to  the associated spin-up channel as evident from Fig.~\ref{fig1}d. 
This  indicates that the increase in the yield observed in Figs.~\ref{fig1}a-c predominantly 
originates from the spin-down channel.
At variance, the decrease in harmonic yield in higher energies matches well for both the  
spin-channels.  This is a strong indication that the defect states do not play a direct role in this 
part of the HHG spectrum of $V_B$, but  instead that
bulk bands predominantly affect this spectral region, as  discussed below.

\subsection{Gap states and electron dynamics}

\begin{figure}[h!]
\includegraphics[width=13 cm]{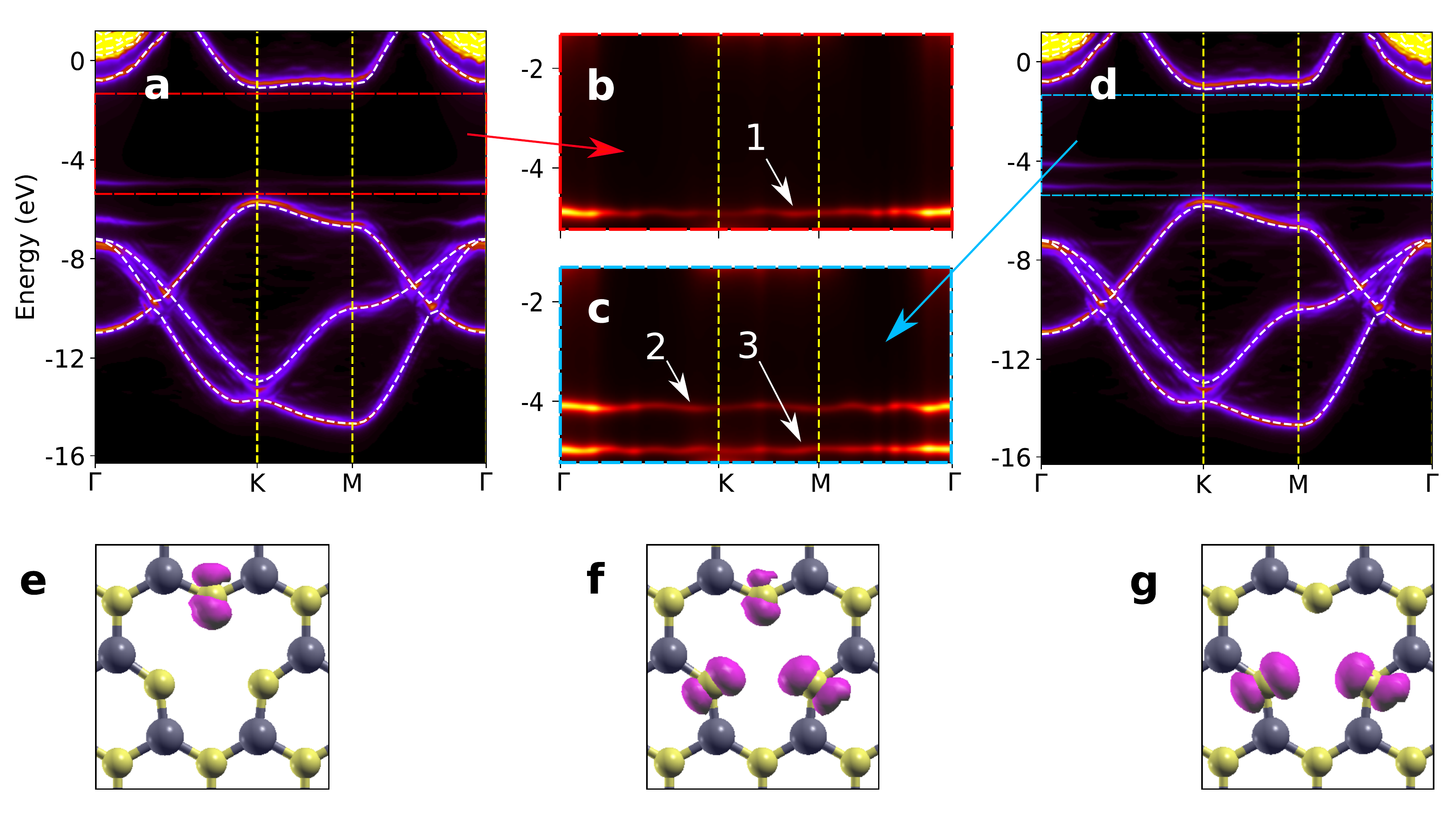}
\caption{{ \bf Spin-resolved unfolded energy band structure and wavefunctions of boron-vacant h-BN ($V_{B}$).}  
Spectral function for {\bf{(a)}} spin-up and {\bf{(d)}} spin-down components of $V_{B}$.  
The pristine band-structure is plotted with white dotted lines for reference. The in-gap portion of the spectral function for spin-up and spin-down components are zoomed respectively in {\bf{(b)}} and {\bf{(c)}}, where vacancy states are recognised  by 1 (for spin-up); and by 2 and 3 
 (for spin-down).
 {\bf{(e)}-(g)} The absolute wavefunctions of the in-gap vacancy states as 
 recognised in {\bf{(b)}} and {\bf{(c)}}, respectively.} 
 \label{band_boron}
\end{figure}

To understand the difference in the harmonics yield associated with spin-up and spin-down channels, let us analyse the ground-state energy band-structure.
The unfolded band-structures of $V_B$ for spin-up and spin-down channels are shown in Figs.~\ref{band_boron}a and  ~\ref{band_boron}d, respectively. 
The band-structure within the band-gap region is zoomed for both the channels and shown in Figs.~\ref{band_boron}b and  ~\ref{band_boron}c.  
As visible from the figure, 
there is one spin-up  (labeled  as 1 in Fig.~\ref{band_boron}b) and two spin-down  (labeled  as 2 and 3 in Fig.~\ref{band_boron}c) 
defect levels within the band-gap of pristine h-BN.   One defect state corresponding to the spin-up channel is pushed within the valence bands (see Fig.~\ref{band_boron}a). 
All three defect states within the band-gap are found to be unoccupied. These factors make the $V_B$ a triple acceptor.  
The corresponding wavefunctions for these defect states are presented 
in Figs.~\ref{band_boron}e-g. The wavefunctions are found to be localised around the vacancy
as expected for dispersionless states.
The $p_{x}$ and $p_{y}$ states of the nitrogen atoms in the vicinity of a boron vacancy 
contribute to these defect states, giving a  $\sigma$-character to the vacancy 
wavefunctions~\cite{liu2007ab}.

Unlike the pristine h-BN, spin-up and spin-down electrons in $V_{B}$ see a
different band-structure near the band-gap, since the spin-resolved in-gap states are different. 
Therefore, spin-up and spin-down electrons evolve differently in the presence of the laser 
pulse.
It means that interband transitions and ionisation involving the defect states will contribute 
differently to the spectrum. 
Hence, the spectral enhancement of the third harmonic can be understood as follows:
As visible from the spin-down band-structure, there is an  additional defect state near the 
valence band, which allows spin-down electrons to be ionised or to recombine
through multiple channels and contribute more to the third harmonic (see Fig.~\ref{fig1}). 

\begin{figure}[h!]
\includegraphics[width=17 cm]{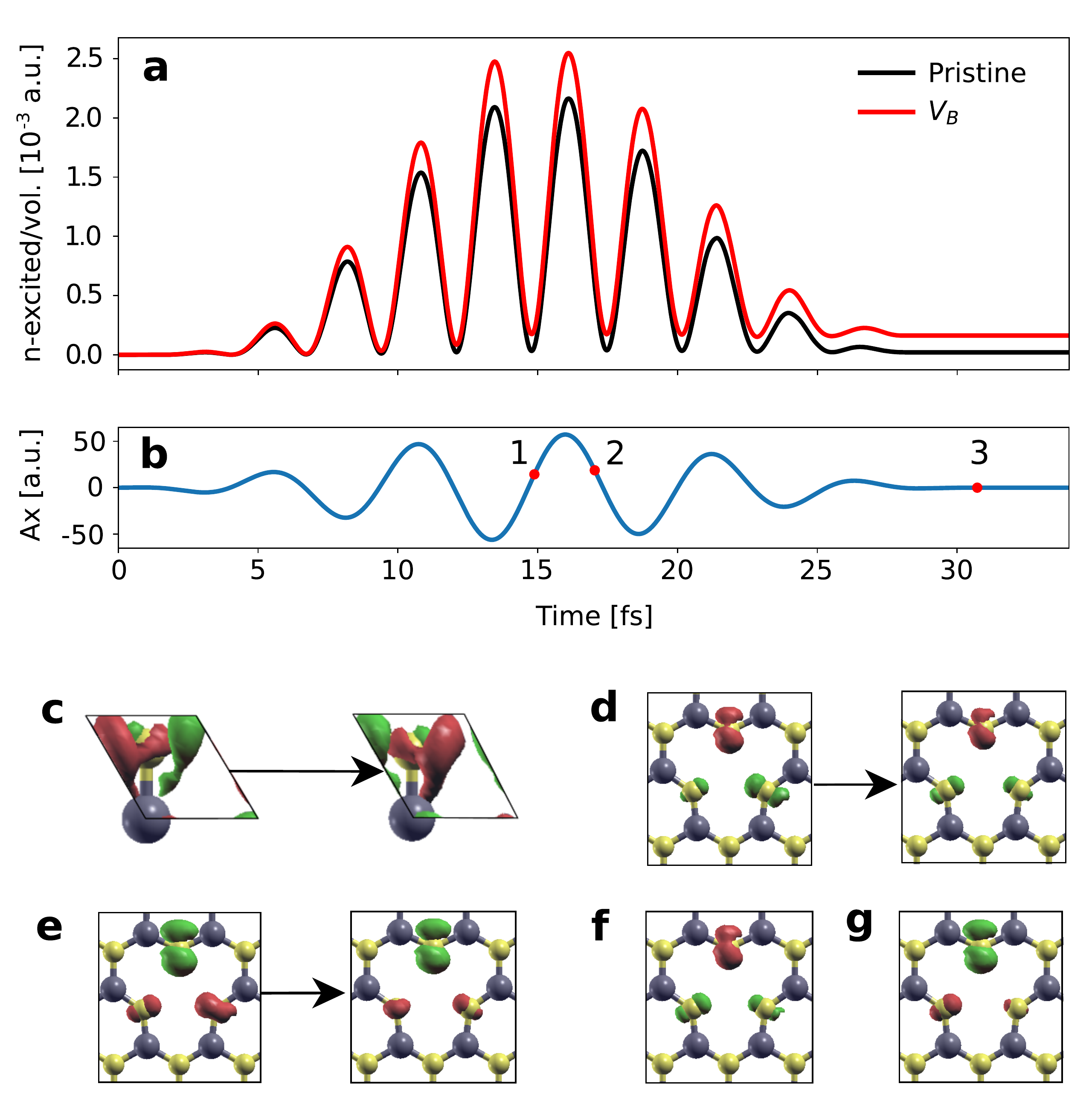}
\caption{{ \bf Time-dependent electron dynamics in boron-vacant h-BN ($V_{B}$).} 
 {\bf{(a)}} Number of excited electrons per unit volume in pristine h-BN (black colour) and $V_{B}$ (red colour), and {\bf{(b)}} vector potential of the driving laser pulse.
Snap shots of the time-evolving induced electron density ($n_{ind}$) near the peak of the vector potential [marked as 1 and 2 in {\bf{(b)}}] for {\bf{(c)}}  pristine h-BN, {\bf{(d)}}  spin-up, and {\bf{(e)}}  spin-down channels 
in $V_{B}$. $n_{ind}$ for {\bf{(f)}}  spin-up and {\bf{(g)}}  spin-down channels 
in $V_{B}$ after the end of the laser pulse [marked as 3 in {\bf{(b)}}]. Green and red colours in the induced electron density stand for positive and negative values, respectively. } 
\label{density_vb}
\end{figure}

It is evident from the band structure that additional defect-states 
effectively reduce the minimal band gap needed to reach the conduction bands. 
Due to the relaxation of atoms next to the vacancy, 
the pristine bands are also slightly modified. However, this modification 
is found  to be negligible compared to the photon energy of the laser and is not further 
discussed.
In order to understand how the presence of the defect states influences the interband-tunneling,  the number of excited electrons during the laser pulse is  evaluated (see 
Fig.~\ref{density_vb}a).
In the presence of defect states, there are mostly two possible ways in which ionisation can be 
modified compared to the bulk material. 
A first possibility is the direct ionisation of the defect states if they are occupied, 
or filling them if they are originally unoccupied. Another possibility is a double sequential 
ionisation, in which the defect states play an intermediate role in easing the ionisation to the 
conduction bands.
In $V_{B}$, there is a finite probability of finding the electrons  in conduction 
bands even after the laser pulse is over (see the red curve  in Fig.~\ref{density_vb}a). 
In contrast to this,  the pulse is not able to promote a significant portion of the valence electrons 
to the conduction bands permanently in the case of pristine h-BN as the band-gap
is significantly large (see the black curve in Fig.~\ref{density_vb}a). 
More precisely, for the $5\times 5$ supercell, we found that 1.6 electrons are ionised, 
compared to 0.25 in the case of the same cell in the pristine material. \\
To achieve a better understanding of the possible ionisation mechanisms, we also consider the 
induced electron density ($n_{ind}$) at two different times near the peak of the vector potential  
(marked as 1 and 2 in Fig.~\ref{density_vb}b) and at the end of the laser field 
(Figs.~\ref{density_vb}f-g).
As reflected in Figs.~\ref{density_vb}d and ~\ref{density_vb}e,
the spin-polarised induced densities of $V_B$  have 
a pronounced localised component near the defects and 
resemble the spatial structure of the initially unoccupied defect wavefunctions (see 
Figs.~\ref{band_boron}e-g). 
The induced densities at the end of the vector potential show that electrons remain in the 
defect states even after the laser pulse is over (see Figs.~\ref{density_vb}f and ~\ref{density_vb}g). 
Considering that the three defect states are originally unoccupied, we cannot conclude that 
more electrons are ionised to the conduction bands. It is most likely that in-gap defect states 
are filled during the laser excitation as ~1.6 electrons are excited and $V_B$ is a triple 
acceptor.

Overall, these results show that the electron dynamics in acceptor-doped solids imply 
a net transfer of population to the originally unoccupied gap states, but for a wide bandgap host 
no more electrons are promoted to conduction bands. This explains why only the low-order 
third harmonic is directly affected by the presence of spin-polarised defect states (as evidenced 
by the spin dependence of the spectrum). 
The low density of these defect states means that fewer photons are absorbed. 
We note that the irreversible population change, assisted by the defect states, ultimately 
implies that more energy is absorbed by the defected solid, which leads to a lower damage 
threshold in comparison to the pristine. However, the intensity considered here is low enough 
to see such an effect.

\subsection{Effect of electron-electron interaction}

We have established so far that the increase of the low-order harmonic yield is compatible with 
the presence of the defect states in the band gap of h-BN. This is an independent-particle 
vision, in which we used the ground-state band-structure of $V_{B}$ to explain the observed 
effect on the HHG spectrum.  We now turn our attention to the higher energy harmonics, for 
which the harmonic yield is decreased. This seems not to be compatible with a simple vision in 
terms of the single-particle band structure, especially in view of the fact that more electrons are 
excited by the laser pulse, as shown in Fig.~\ref{density_vb}a. \\
To understand this, let us investigate  the effect of the electron-electron interaction on the 
electron dynamics in $V_{B}$  and pristine h-BN. 
Within the dipole approximation, the HHG spectrum can be expressed as 
~\cite{tancogne2017impact, stefanucci2013nonequilibrium} 
\begin{equation}
\textrm{HHG}(\omega)  
\propto  \left| \mathcal{FT} \left[ \int d\textbf{r} ~\left\lbrace n_{ind}(\textbf{r},t) + n_0(\textbf{r}) 
\right\rbrace \nabla v_0 (\textbf{r}) \right] + N_e \textbf{E}(\omega)  \right|^2.
\label{current}
\end{equation}
Here, $n_{ind}(\textbf{r},t)$ is the induced electron density, 
$v_0$ is the electron-nuclei interaction potential, $N_e$ is the total number of electrons, and $
\textbf{E}$ is the applied electric field. $n_{ind}(\textbf{r},t)$   is the difference of  the total 
time-dependent electron density $n(\textbf{r},t)$ and the ground-state electron density 
$n_0(\textbf{r})$, i.e., $n_{ind}(\textbf{r},t) = n(\textbf{r},t) - n_0(\textbf{r})$.  Also, $n(\textbf{r},t)$ 
is  decomposed in spin-polarised fashion as  $n(\textbf{r},t)$ = $n_{\uparrow}(\textbf{r},t)$ + 
$n_{\downarrow}(\textbf{r},t)$.
If one analyses this expression, it is straightforward to understand how the introduction of  
vacancies can change the harmonic spectrum, through a change in the local potential structure 
near the defect.  
This results in the change in gradient of the electron-nuclei interaction potential $v_0$, which is 
independent of electrons interacting with each other or not. 
Apart from this explicit source of change, it is clear that the dynamics of the induced density, 
evolving from a different ground-state also leads to the modifications in 
the harmonic spectra. The fact that the ground states are different, and hence $n_0(\textbf{r})$ 
is different, does not affect the harmonic spectrum because of the absence of time dependence 
in both $n_0$ and $v_0$. The possible difference between the HHG spectra can, therefore, 
be understood in terms of independent particle effects (originating from the structural change of 
the nuclear potential $v_0$) and interaction effects from the induced density $n_{ind}$.

\begin{figure}[h!]
\includegraphics[width=10 cm]{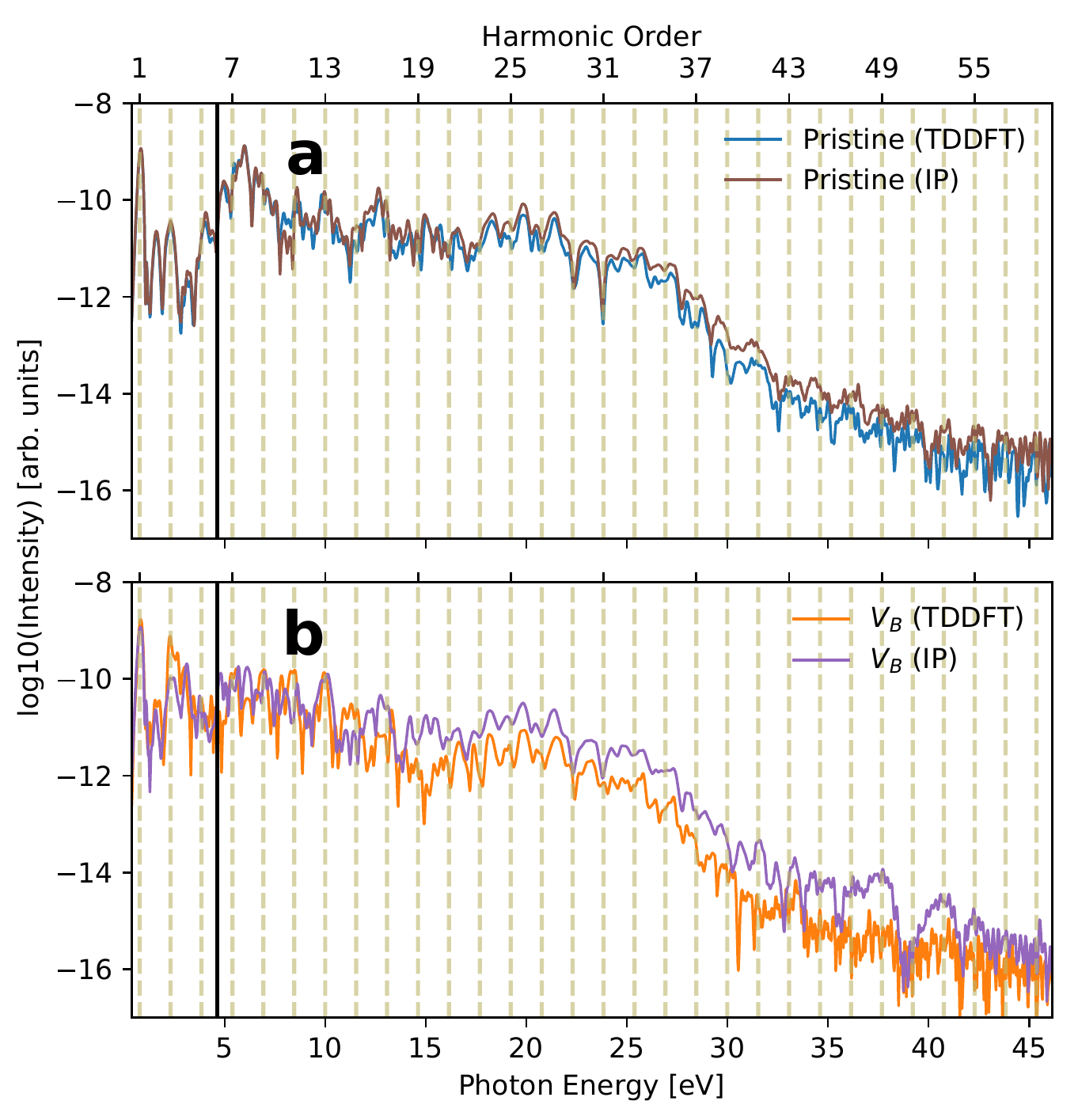}
\caption{{\bf Many-body effect in harmonic spectrum}. {\bf{(a)}} HHG spectrum for pristine h-BN calculated using TDDFT (blue) and the independent particle (IP) approximation (brown). {\bf{(b)}} HHG spectrum for $V_B$ calculated using TDDFT (orange) and IP approximation (violet). The black vertical line represents the energy band gap of pristine h-BN.} 
\label{corr}
\end{figure}

In order to disentangle these two sources of differences between pristine and defect h-BN, we 
simulate  the harmonic spectra 
within an independent particle (IP) model by freezing the Hartree and exchange-correlation 
potentials to the ground-state value. The harmonic spectra for pristine h-BN and $V_B$ within 
TDDFT and IP are compared in Fig.~\ref{corr}.  
In the case of pristine h-BN, the HHG-spectra  obtained by the IP model and 
TDDFT are similar. Hence, there is no significant many-body effect in HHG from pristine h-BN 
with an in-plane laser polarisation, at least as described by the PBE functional used here. 
A similar finding has been reported for Si~\cite{tancogne2017impact} and 
MgO~\cite{tancogne2017ellipticity}, within the local-density approximation.
Only the high-order harmonics display small differences and there the electron-electron 
interaction reduces  the harmonic yield, as found in Ref.~\cite{yu2019enhanced}.
In contrast, the HHG-spectra obtained by IP and 
TDDFT are  significantly different for $V_{B}$ as reflected in Fig.~\ref{corr}b.  
This indicates that the electron-electron interaction is essential for HHG in defected solids. 
The total currents corresponding  to the harmonic 
spectra for the pristine h-BN and $V_B$  are shown in Supplementary Fig.~2.  \\
For the case of $V_B$, the induced electron density is displayed in Fig.~\ref{density_vb}. This 
helps us to understand how the spatial structure of the defect wavefunctions influences the  
spectrum.
Similarly, Fig.~\ref{density_vb}c indicates that the spatial density oscillations are 
responsible for HHG in pristine h-BN. It is clear that the induced density is different in the two 
systems.
The substantial  difference in the harmonic spectra of $V_{B}$ obtained by two approaches, 
TDDFT and IP,  is due to the so-called local field effects. This  is explored in detail in the 
following paragraphs. It will be shown that this difference is responsible for the decrease of the 
harmonic yield for $V_B$.

In the presence of an external electric-field, the localised induced-charge  acts as 
an oscillating dipole near the vacancy. The dipole induces a local electric field, which 
screens the effect of the external electric field. This is usually referred to as local field effects. 
The same mechanism is responsible for the appearance of a depolarisation field at the surface 
of a material driven by an out-of-plane electric field~\cite{le2018high, tancogne2018atomic}. It 
is important to stress that this induced dipole is expected to play a significant role here, due to 
the $\sigma$-character of the vacancy wavefunctions.
The induced electric field is directly related to the electron-electron interaction term 
as clearly shown in Ref.~\cite{le2018high}. 
As shown in Fig.~\ref{corr}b, the harmonic yield at higher energies is increased
if we neglect local field effects, i.e., we treat electrons at the IP level. 
Moreover, the electron-electron interaction also affects the third harmonic, as shown in 
Fig.~\ref{corr}, but lead to an increase of the harmonic yield. We, therefore, attribute this effect 
not to local field effects but to correlation effects.
It is important to stress that in Maxwell equations, the source term of the induced electric field 
is the induced  density (summed over spin). 
Therefore, the induced electric field acts equally on 
the HHG from both spin channels, which is why the decrease of the yield occurs equally on 
both spin channels, see Fig.~\ref{fig1}b. 

We conclude that the modifications in the HHG spectrum due to a point defect originate from a 
complex interplay of two important factors: the electronic transitions including the in-gap defect 
states, and the electron-electron interaction.  
This indicates that HHG in defected-solids  cannot be fully addressed via 
IP approximation as this can lead to  wrong qualitative predictions in certain cases. 

Finally, we discuss the dependence of the defect concentration on the HHG spectrum by 
computing the HHG for  a 7$\times$7 supercell with a boron vacancy, which corresponds to a $
\sim$1\% doping concentration. The harmonic spectrum is presented in Supplementary Fig.~4. 
The third harmonic enhancement persists with comparable intensity even where there is a 
lower defect
concentration, whereas the higher energy region of the harmonic spectrum matches well with 
the
pristine spectrum. This is consistent with the observation  that the
higher energy spectrum is dominated by the bulk bands. 
This indicates that some of the present findings depend on the defect concentration and may 
not be observed below a certain concentration threshold.

\subsection{High-harmonic generation in hexagonal boron nitride with a nitrogen vacancy}

\begin{figure}[h!]
\includegraphics[width=13 cm]{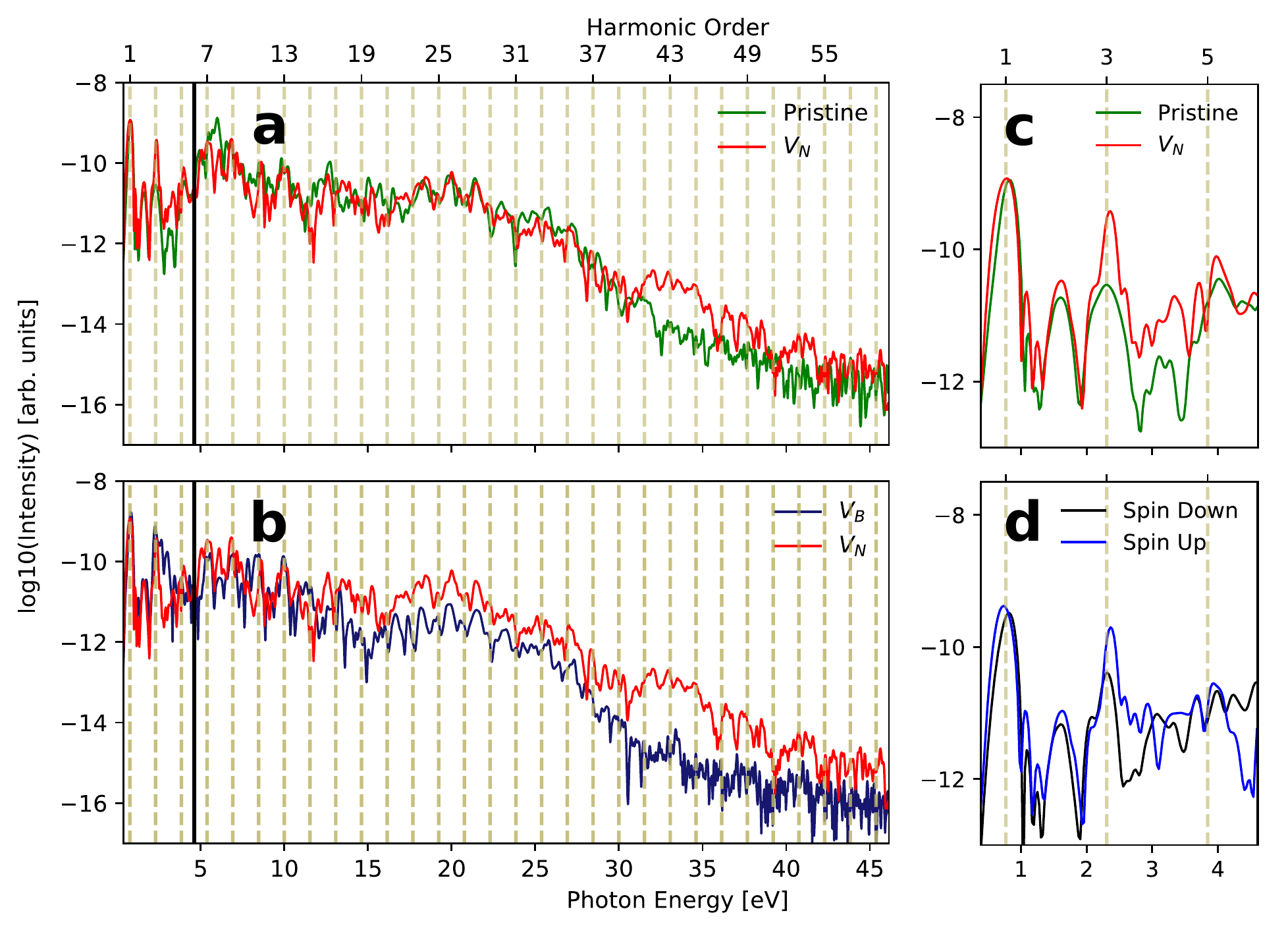}
\caption{
{\bf High-order harmonic spectrum of h-BN with nitrogen vacancy}.  
Total harmonic spectrum of {\bf{(a)}}  nitrogen vacant h-BN ($V_{N}$) and pristine h-BN,  
{\bf{(b)}} nitrogen-vacant h-BN ($V_{N}$) and boron-vacant h-BN ($V_{B}$), 
and {\bf{(c)}}  below band-gap harmonic spectrum of  $V_N$ and pristine h-BN.  
The black vertical line represents the energy band-gap of pristine h-BN.
Spin-resolved harmonic spectrum for {\bf{(d)}}  spin-up and {\bf{(e)}}  spin-down components,  and its comparison with the total below band-gap harmonic spectrum of $V_N$.} 
\label{hhg_vn}
\end{figure}

Until now, we have discussed HHG in h-BN with a boron vacancy. 
Now we will explore HHG in h-BN with a nitrogen vacancy.
Similar to the $V_B$ case, h-BN with a nitrogen vacancy ($V_N$) is also spin polarised.
Fig.~\ref{hhg_vn}a presents the high-order harmonic spectrum 
of $V_{N}$ and 
its comparison with the spectrum of pristine h-BN. 
As apparent from the figure, the spectrum  of $V_{N}$ 
resembles the pristine spectrum more closely, except in the below band-gap regime and for an 
increase of the yield between 30 to 35 eV.
All the laser parameters are identical to  the case of $V_{B}$. 
The spectra of $V_{N}$ and $V_{B}$ could be expected  to resemble each other as one atom 
from the pristine h-BN has been removed in both of them. 
However, this is not the case, as evident from Fig.~\ref{hhg_vn}b and we note that the high-energy part of the spectrum of $V_N$ is much closer to the one of pristine h-BN than $V_B$, 
except above 30 eV.  The spectra of 
$V_{N}$ and $V_{B}$ are also fairly different. 
On close inspection of the  below band-gap spectrum, one finds that the 
third harmonic in  $V_{N}$ is significantly enhanced  with respect to  the pristine case
(see Fig.~\ref{hhg_vn}c). The same observation was made in the case of HHG in $V_{B}$. 
To understand whether the reason behind this enhancement is identical to $V_{B}$ or not, we 
examined the spin-resolved harmonics in $V_{N}$. 
The below band-gap spin-resolved spectrum 
reveals that the third harmonic has a greater contribution from the spin-up electron than the 
spin-down electron (see Fig.~\ref{hhg_vn}d; also see Supplementary Fig.~5 for the 
corresponding spectra in the full energy range). This contrasts the findings in $V_{B}$ where 
the major contribution originated from the spin-down electron. 
To understand this behaviour, let us explore the ground-state band-structure of $V_{N}$. 

\begin{figure}[h!]
\includegraphics[width=15 cm]{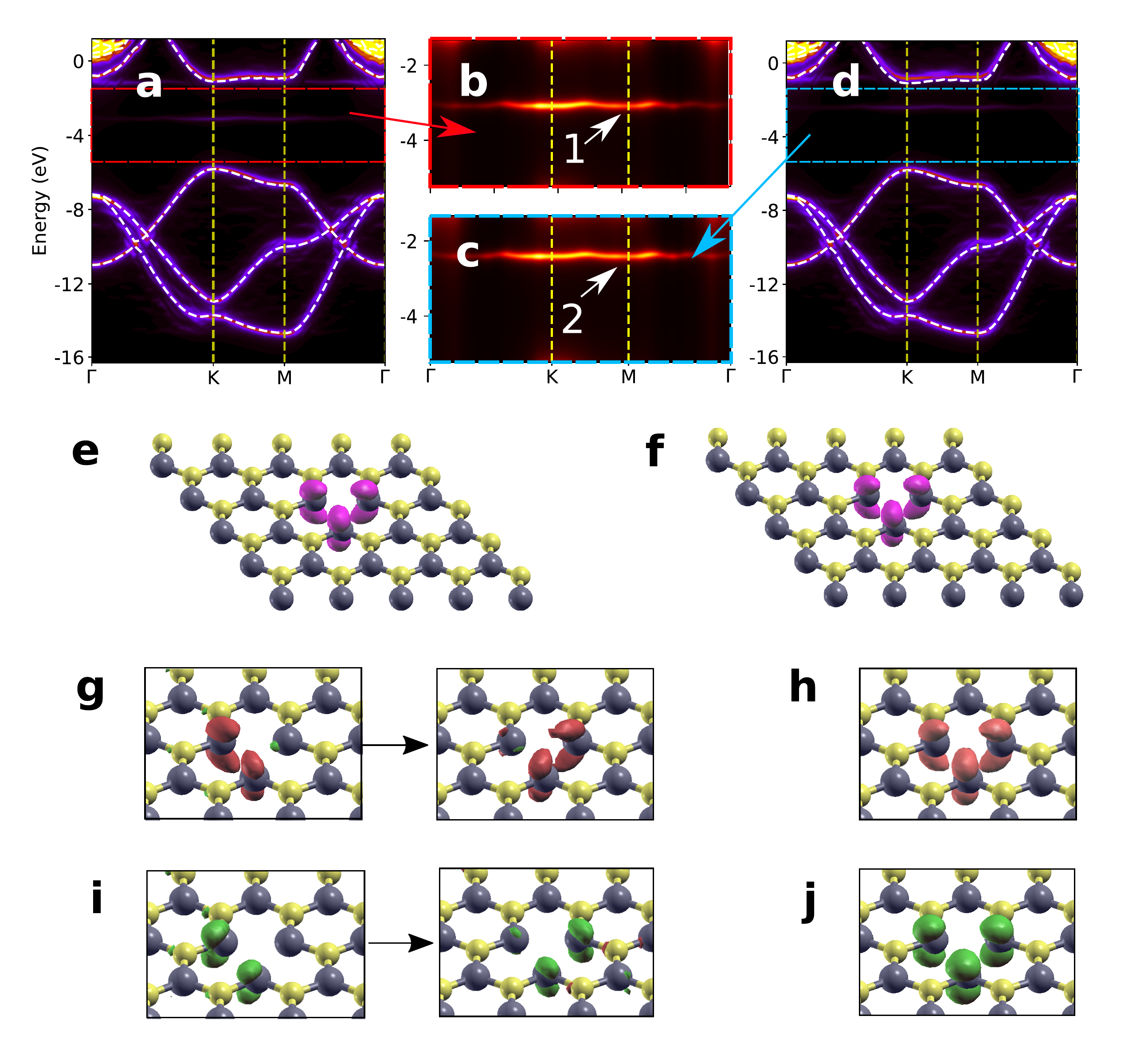}
\caption{
{ \bf Spin-resolved unfolded band structure, absolute wavefunctions and induced electron density of nitrogen-vacant h-BN ($V_{N}$).}  
Spectral function for {\bf{(a)}} spin-up and {\bf{(d)}} spin-down components of $V_{N}$. 
The pristine band-structure is plotted with white dashed lines for reference. The in-gap portion of the spectral function for spin-up and spin-down components are zoomed respectively in (\textbf{b}) and (\textbf{c})  in which 
the in-gap vacancy states are recognised  by 1 (for spin-up); and by 2 
 (for spin-down). 
 {\bf{(e)}-(f)} The absolute wavefunctions of the in-gap vacancy states 
 1 and 2 in {\bf{(a)}} and {\bf{(b)}}, respectively.
 Snap shots of the time-evolving induced electron density ($n_{ind}$) near the peak of the vector potential for {\bf{(g)}} spin-up and {\bf{(i)}} spin-down channels; and 
after the end of the vector potential  for {\bf{(h)}} spin-up and {\bf{(j)}} spin-down channels 
in $V_{N}$. The  vector potential is shown in Fig. ~3{\bf{b}}. Green and red colours in the induced electron density stand for positive and negative values, respectively.} 
\label{band_nitrogen}
\end{figure}

The unfolded band-structure of $V_{N}$ is presented in 
Fig.~\ref{band_nitrogen}. 
Each of the three boron atoms has an unpaired electron following the removal of a nitrogen 
atom 
from pristine h-BN.   
One spin-up and one spin-down vacancy states emerge within the band-gap and are located 
more closely to the conduction band (see Figs.~\ref{band_nitrogen} a-d). 
One additional defect state  is pushed further towards the conduction band in each case. The 
spin-up defect state is found to be occupied. This makes $V_N$ a single donor 
\cite{huang2012defect}. The $p_{z}$-states of the boron atoms, which are in the vicinity of a 
nitrogen vacancy,  contribute to the defect states.  
This gives a $\pi$-character to the defect wavefunctions  (Figs.~\ref{band_nitrogen}e and 
\ref{band_nitrogen}f)~\cite{huang2012defect,azevedo2009electronic}.  
Similar to $V_{B}$, only gap states are analysed. 
The spin-up defect level is 
occupied and close to the conduction band, which explains 
the  major contribution of the spin-up electron to the third harmonic in $V_{N}$.  
Electrons in this state can easily get ionised to 
the conduction bands and add more spectral weight  to the third harmonic. 
Note that, unlike $V_{B}$, the local symmetry in $V_{N}$ is preserved, which also explains why 
the HHG spectrum from $V_N$ is close to the HHG from pristine h-BN. 

The unfolded band-structures of  $V_{N}$ and $V_{B}$ mainly
explain the significant enhancement of the third harmonic and its different spin-polarised 
nature.
To explain the overall difference in the harmonic spectrum,  
the snap-shots of $n_{ind}$ in $V_{N}$ at different times along the vector potential are 
presented in Figs.~\ref{band_nitrogen}g-j. 
In the real-space picture, defect contribution arises from the localised induced density, which
can be seen from the integral in Eq.~(\ref{current}). 
The depletion of the spin-up defect state as well as induced electron density in the spin-down 
defect state can be observed at the end of the pulse (see Figs.~\ref{band_nitrogen}h and 
\ref{band_nitrogen}j).

In comparison to $V_B$, the effect of screening due to the local field effects is weaker in the 
case of $V_N$. 
In comparison to $V_B$, 
the weaker impact of the local field effects on $V_N$ can be attributed to the following two 
reasons: 1) the induced charge density has a pronounced localised component around the 
nitrogen vacancy. As evident from Figs.~\ref{band_nitrogen}e and  ~\ref{band_nitrogen}f, the 
wavefunctions of spin-up and spin-down electrons have similar spatial structure, whereas the 
corresponding induced charge densities have opposite sign (see Figs.~\ref{band_nitrogen}g-j). 
This partial cancellation of the spin-resolved induced charge density lowers the total induced 
charge and results in weaker local field effects. 2) For the in-plane laser polarisation, the 
induced dipole due to the $\pi$-like defect states gets much less polarised than for the  $
\sigma$-like defect states. This results in  weaker screening in $V_{N}$ than $V_{B}$.

The  weaker local field effects in $V_N$ are fully consistent with the HHG spectra of $V_N$, 
obtained through TDDFT and IP models (see Supplementary Fig.~3). Here, the third harmonic 
enhancement is well captured within the IP model, but the increase in the yield between 30 to 
35\,eV is found to 
originate from the electron-electron interaction (see Supplementary Fig.~3). Finally, in the 
HHG spectrum of $V_N$,  the defect state plays a role even at higher orders (see the spin-resolved spectra in Supplementary Fig.~5), though these effects are feeble.

In essence, the total HHG spectra corresponding to  $V_B$ and $V_N$ include  the 
contributions  from the energy-bands of  
pristine h-BN as well as from the transitions including gap states. 
Moreover,  electron-electron interaction plays significant and different roles 
in both cases.
The harmonic spectrum of the defected solid preserves some piece of information of the 
pristine structure in the higher energy regime 
along with the characteristic signatures of the defect in the near band-gap regime, or close to 
the cutoff energy for $V_N$.

\section{Discussion}

In summary, we have investigated the role of vacancy-defect in solid-state HHG. 
For this purpose, we considered h-BN with a boron or a nitrogen atom vacancy. 
In simple terms, 
one may assume  that h-BN with a boron atom vacancy 
or with a nitrogen atom vacancy would exhibit similar HHG spectra since a 
single atom from h-BN has been removed. However, this is not the case as boron and 
nitrogen vacancies lead to qualitatively different electronic structures, and this is apparent from 
their corresponding 
gap states. It has been found that once an atom is removed from h-BN, either boron or 
nitrogen, the system becomes spin-polarised with a non-zero magnetic moment near the 
vacancy. As a consequence, the defect-induced gap states are found to be different for each 
spin channel and for each vacancy, which we found to be strongly reflected in the low-order 
harmonics.  These contributions are strongly spin-dependent, according to the ordering and 
occupancy of the defect states. Altogether, the role of the defect states can be understood 
by analysing the spin-polarised spectra, and the findings are in accordance with the spin 
polarised band-structure. This establishes one aspect of the role of defect states in strong-field 
dynamics in solids.

In addition, the vacancy wavefunctions of $V_{B}$ and $V_{N}$ show $\sigma$- and $\pi$-characters respectively, which lead to different qualitative changes in the harmonic spectra of 
vacancies, due to the local-field effects and electron correlations. 
These different behaviours are caused by the creation or absence of an induced dipole, which 
may counteract the driving electric field, and directly depends on the spatial shape of the defect 
state wavefunctions.
Moreover, the electron-electron interaction also manifests itself in the decrease 
of the harmonic yield close to the energy cutoff in the $V_B$ case, whereas this effect is 
completely absent in the $V_N$ or pristine cases.
This implies that the nature of the vacancies in $V_{B}$ and $V_{N}$ is entirely different, as 
reflected in their HHG spectra, even at a defect concentration as low as 2\%. 
The HHG spectrum of $V_{N}$ is similar to the pristine h-BN, 
whereas the spectrum of $V_{B}$ differs significantly. These effects essentially imply that some 
defects are more suited to the modification of the HHG spectra of the bulk materials. This in 
turn opens the door for tuning HHG by engineering defects in solids.

From our work, we can also estimate of other known defects in h-BN.
If one considers the doping impurity instead of vacancy, e.g., carbon impurity, the band-
structure of h-BN remains spin-polarised in nature near the band 
gap~\cite{attaccalite2011coupling,huang2012defect}.  If a boron atom is replaced by a carbon 
atom, one occupied spin-up and one unoccupied spin-down defect levels appear. Both 
the defect levels are near the conduction band with the wavefunctions contributed from the 
$p_{z}$ orbitals of carbon and nearby nitrogen atoms. So, the carbon doping defect is 
expected to show qualitatively similar behaviour in the HHG spectra as that observed in the 
case of $V_{N}$. On the other hand, one occupied spin-up and one unoccupied spin-down 
states appear near the conduction bands when nitrogen is replaced  with a carbon atom.  The 
wavefunction in this case is contributed mostly by the $p_{z}$ orbitals of the carbon as well as 
nearby boron atoms. In this case, the  enhancement in the below band-gap harmonics  is 
expected due to the defect states near the valence bands similar to $V_{B}$. However, the 
effect of screening is expected to be lower compared to $V_{B}$ as the nature of 
wavefunctions here is similar to that of $ V_{N}$.\\
In the case of bi-vacancy in h-BN, one occupied defect state exists near the valence band and 
two unoccupied defect states near the conduction band (see 
Ref.~\cite{attaccalite2011coupling}). In this situation, if the separation between defect states 
and the nearby bands is small compared to the photon energy,  
no significant changes in the below band-gap harmonics are expected.\\
Let us finally comment on the appealing possibility of performing imaging of spin-polarised 
defects in solids using HHG. As spin channels are not equivalent in the studied defects, one might think about using circularly polarised pulses to probe each spin channel independently.
However, in the context of dilute magnetic impurities, as studied here, a crystal will host as 
many defects with positive magnetic moments  as the negative ones, and the signals for up or down spin channel will appear  as identical after macroscopic averaging. \\
Our work opens up interesting perspectives for further studies on strong-field electron dynamics in two-dimensional and extended systems, especially involving isolated defects. 
Further work may address the possibility of monitoring electron-impurity scattering using HHG, more complex defects such as bi-vacancies, and a practical scheme for imaging buried defects in solids.

\section{Methods}

\subsection{Geometry relaxation}

Geometry optimisation was performed using the DFT code Quantum ESPRESSO \citep{giannozzi2009p,giannozzi2017p}. 
Both the atomic coordinate and the  lattice constant relaxation were allowed. 
Forces were optimised to be below 10$^{-3}$ eV/\AA. We used an energy cutoff of 150 Ry., and a k-point grid of 10$\times$10$\times$1  $\mathbf{k}$-points. We used a vacuum region of more than 20 \AA~ to isolate the monolayer from its periodic copies.
The relaxed lattice constants of 12.60 and 12.57 \AA~ for $V_B$, and 12.48 \AA~ for $V_N$ were obtained while for pristine h-BN, it was found to be 12.56 \AA.
The structure of h-BN with a boron vacancy was relaxed by  lowering the local threefold symmetry~\cite{huang2012defect}. The lowering of the symmetry with the boron vacancy is attributed to the Jahn-Teller distortion, which is found to be independent of the defect concentration~\cite{huang2012defect,liu2007ab,attaccalite2011coupling}.

\subsection{TDDFT simulations}
By propagating the Kohn-Sham equations within TDDFT, the evolution 
of the time-dependent current is computed by using the 
Octopus package~\cite{andrade2015real, castro2004propagators}. 
TDDFT  within generalised gradient approximation (GGA) with exchange and correlations of Perdew-Burke-Ernzerhof (PBE)~\cite{perdew1996generalized} is used for all the simulations presented here. The adiabatic approximation is used for all the time-dependent simulations. 
We used norm-conserving pseudopotentials. The real-space cell was sampled with a grid spacing of 0.18 \AA, and  a 6$\times$6 $\mathbf{k}$-points grid was used to sample the 2D Brillouin zone. The semi-periodic boundary conditions are employed.  A simulation box of 74.08 \AA~ along the nonperiodic dimension, which includes 21.17 \AA~of absorbing regions on each side of the monolayer, is used. The absorbing boundaries are treated using the complex absorbing potential method and the cap height h is taken as h = -1 atomic units (a.u.) to avoid the reflection error in the spectral region of interest~\cite{de2015modeling}.

Effective single-particle band-structure or spectral function for the vacancy structures is visualised by unfolding the band-structure of 5$\times$5  supercell with 50 atoms ~\citep{ku2010unfolding, popescu2012extracting, medeiros2014effects}.   
The spin-polarised calculations are used to address the spin-polarised  vacancies. 

The Fourier-transform of the total spin-polarised 
time-dependent electronic current $\textbf{j}_{\sigma}(\textbf{r},t)$ is used to 
simulate the  HHG spectrum as  
\begin{equation}
\textrm{HHG}(\omega) = \left| \mathcal{FT} \left[ \sum_{\sigma = \uparrow,\downarrow} 
\frac{\partial}{\partial t}
\left( \int d \textbf{r} ~\textbf{j}_{\sigma}(\textbf{r},t)\right)  \right] \right|^2,
\end{equation}
where $\mathcal{FT}$ and $\sigma$ stand for the Fourier-transform and the spin-index, respectively.  

The total number of excited electrons is obtained by projecting the time-evolved wavefunctions ($|\psi_n(t)\rangle$) on the basis of the ground-state wavefunctions ($|\psi_{n'}^{\mathrm{GS}}\rangle$) 
\begin{equation}
 N_{\mathrm{ex}}(t) = N_e - \frac{1}{N_\mathbf{k}}\sum_{n,n'}^{\mathrm{occ.}}\sum_{\mathbf{k}}^{\mathrm{BZ}} |\langle \psi_{n,\mathbf{k}}(t) | \psi_{n',\mathbf{k}}^{\mathrm{GS}} \rangle|^2,
\end{equation}
where $N_e$ is the total number of electrons in the system and $N_{\mathbf{k}}$ is the total number of  $\mathbf{k}$-points used to sample the BZ. The sum over the band indices $n$ and $n'$ run over all occupied states.

\section*{Code Availability}
The OCTOPUS code is available from http://www.octopus-code.org.

\section*{Data Availability}
The data that support the findings of this study are available from the corresponding authors upon request, and will be deposited on the NoMaD repository.

 \section*{Competing financial interests}
 The Authors declare no Competing Financial or Non-Financial Interests.

\section*{Acknowledgments}

This work was supported by the European Research Council (ERC-2015-AdG694097), the Cluster of Excellence (AIM), Grupos Consolidados (IT1249-19),  SFB925, the Flatiron Institute  (a division of the Simons Foundation), and  Ramanujan fellowship (SB/S2/ RJN-152/2015).

\section*{Contributions}
N. T.-D., A. R. and G. D.  conceived the idea, designed the research and supervise the work. M. M. S. performed all the calculations. All authors discussed the results and contributed to the final manuscript.
\newpage

\pagebreak 

\end{document}